\documentclass[12pt]{article} 
\usepackage[top=31mm, bottom=46mm, left=29mm, right=29mm]{geometry}

\usepackage{mathtools}
\numberwithin{equation}{section}

\usepackage{setspace} 

\usepackage{indentfirst}
\usepackage{cite} 					

\usepackage{hyperref}
\hypersetup{
	unicode,	
	colorlinks,
	citecolor=[rgb]{0,.3,.5}, 
	linkcolor=[rgb]{0,.3,.5},
	urlcolor=[rgb]{0,.3,.5}, 
	bookmarksopen=true,
	bookmarksopenlevel=\maxdimen,
} 

\usepackage{amsmath, amssymb} 	
\usepackage{mathrsfs} 			
\usepackage{bm}				

\def\gl#1#2{\ifmmode \mathrm{GL}(#1; {\bf #2}) \else $\mathrm{GL}(#1; {\bf #2})$\fi}
\def\sl#1#2{\ifmmode \mathrm{SL}(#1; {\bf #2}) \else $\mathrm{SL}(#1; {\bf #2})$\fi}
\def\so#1{\ifmmode \mathrm{SO}({#1}) \else $\mathrm{SO}(#1)$\fi}

\def\sp#1#2{\ifmmode \mathrm{Sp}(#1; {\bf #2}) \else $\mathrm{Sp}(#1; {\bf #2})$\fi}
\def\usp#1{\ifmmode \mathrm{USp}(#1) \else $\mathrm{USp}(#1)$\fi}
\def\spin#1{\ifmmode \mathrm{Spin}(#1) \else $\mathrm{Spin}(#1)$\fi} 
\def\su#1{\ifmmode \mathrm{SU}({#1}) \else $\mathrm{SU}(#1)$\fi}

\def\double #1{#1{\hbox{\kern-2pt $#1$}}}
\def\un#1{\underline #1}

\def\on#1#2{{\buildrel{\mkern2.5mu#1\mkern-2.5mu}\over{#2}}}
\def\dt#1{\on{\hbox{\bf .}}{#1}}                
\mathcode`\*="702A                  
\def\half{{\textstyle{1\over{\raise.1ex\hbox{$\scriptstyle{2}$}}}}}

\begin{document}

\addtolength{\baselineskip}{.5mm}
\thispagestyle{empty}
\begin{flushright}
{
{\sc MI-TH-1617}\\
}
\end{flushright}

\begin{center}
\doublespacing{\Large \textbf {
M-theory Potential from the $G_2$ Hitchin Functional\\ 
in Superspace
}}
\end{center} 
\begin{center} 
{Katrin Becker,\!\footnote{kbecker@physics.tamu.edu}
Melanie Becker,\!\footnote{mbecker@physics.tamu.edu}
Sunny Guha,\!\footnote{sunnyguha@physics.tamu.edu}
William D. Linch III,\!\footnote{wdlinch3@gmail.com}
and\\
 Daniel Robbins\footnote{dgrobbins@albany.edu}}
\\[5mm]
{{}$^{1,2,3,4}$\em
George P. and Cynthia Woods 
Mitchell Institute for \\
Fundamental Physics and Astronomy, \\
Texas A\&{}M University.\\
College Station, TX 77843 USA.
}\\[3mm]
{{}$^5$\em
Department of Physics, \\
University at Albany.\\
1400 Washington Ave.
Albany, NY 12222 USA.
}\\
\end{center} 

\vspace{1mm}

\begin{abstract}
We embed the component fields of eleven-dimensional supergravity into a superspace of the form ${\bm X}\times Y$ where $\bm X$ is the standard 4D, $N=1$ superspace and $Y$ is a smooth 7-manifold.
The eleven-dimensional 3-form gives rise to a tensor hierarchy of superfields gauged by the diffeomorphisms of $Y$. 
It contains a natural candidate for a $G_2$ structure on $Y$, and being a complex of superforms, defines a superspace Chern-Simons invariant.
Adding to this a natural generalization of the Riemannian volume on ${\bm X}\times Y$ and freezing the (superspin-$\frac32$ and 1) supergravity fields on $\bm X$, we obtain an approximation to the eleven-dimensional supergravity action that suffices to compute the scalar potential. 
In this approximation the action is the sum of the superspace Chern-Simons term and a superspace generalization of the Hitchin functional for $Y$ as a $G_2$-structure manifold. 
Integrating out auxiliary fields, we obtain the conditions for unbroken supersymmetry and the scalar potential.
The latter reproduces the Einstein-Hilbert term on $Y$ in a form due to Bryant.
\end{abstract}


\newpage
\tableofcontents

\setcounter{page}1

\section{Introduction}

In the zoo of supergravity theories, eleven-dimensional supergravity is unique in that it has the largest possible (manifest) spacetime symmetry group. 
Despite being, in this sense, the most fundamental of supergravity theories, it has various quite mysterious properties. For example, in contrast to its ten-dimensional relatives, there is no theory of critical superstrings that has it as a low-energy limit. To find a home even somewhat analogous, one must go to M-theory (which is even more mysterious) and take a massless limit of that. 
Another presumably related property is the emergence of an exceptional symmetry of its (gauged) compactifications on tori. 

For applications to the study of physics in lower dimensions, this theory may be compactified on eleven-dimensional manifolds of the form $X\times Y$ and expanded in Kaluza-Klein modes by integrating over $Y$. 
This results in an effective theory on $X$ in which the contribution of the internal part is organized in a tower of ever-more-massive fields. 

An alternative to this approach is to split the eleven-dimensional spacetime as $X\times Y$ and to reorganize the fields into representations of the reduced structure group but without averaging over the ``internal'' space.
Such backgrounds were precisely the subject of 
reference \cite{Becker:2014uya}, wherein this is referred to as ``keeping locality in $Y$''.
There, the action for the bosonic part of eleven-dimensional supergravity was decomposed on $X \times Y$ explicitly. 
Of course it is always possible to keep the full diffeomorphism invariance of the eleven-dimensional theory recast in terms of covariant, interacting $X$ and $Y$ parts.
What is somewhat surprising, however, is that this can be organized in a very manageable form \cite{Becker:2014uya}.
We would like to construct a superspace action that reproduces the bosonic eleven-dimensional supergravity action in this form. 

As this is presumably impossible (in the na\"ive sense) for more than 8 real supercharges, we settle for a superspace description with at most $N=(1,0)$ supersymmetry in 6D, $N=1$ in 5D, or $N=2$ in 4D. 
These maximal off-shell cases require an infinite number of auxiliary fields \cite{Siegel:1981dx} and non-chiral matter. 
This complicates the use of such a superspace description both technically and phenomenologically. 
Instead, we propose to embed the components of eleven-dimensional supergravity into 4D, $N=1$ superfields with arbitrary $Y$-dependence. 
This gives a description of eleven-dimensional supergravity on $\bm X\times Y$ with $\bm X$ a curved superspace modeled on $\mathbf R^{4|4}$ and $Y$ a Riemannian 7-manifold. 
Projecting such a theory to component fields results in a component supergravity description on the bosonic submanifold $X \times Y$.

Although the resulting physics is eleven-dimensionally super-diffeomorphism invariant, only the 4D, $N=1$ part of the local super-Poincar\'e symmetry would be manifest (together with the 7D (bosonic) Riemannian part). 
Note that precisely this amount of local super-Poincar\'e invariance is what one would retain were one to compactify on a manifold $Y$ admitting a Riemannian metric of $G_2$ holonomy. 
Although we do not insist on such a background in this work, it will be useful to adopt the language of 4D, $N=1$ compactifications in which we refer to $X$ or $\bm X$ as ``spacetime'' and $Y$ as the ``internal space''.

There are various partial realizations of this superspace supergravity program less ambitious than the construction of the full 11D theory in arbitrary $\bm X\times Y$.
For example, one could attempt to build the linearized action by working out the linearized superdiffeomorphisms and building an invariant action order-by-order following a superspace Noether procedure.\footnote{In the analogous problem for 5D, $N=1$ supergravity in 4D, $N=1$ superspace, this was the approach taken in \cite{Linch:2002wg, Gates:2003qi}. 
This setup is related to the eleven-dimensional version considered here by taking $Y= \mathbf R \times Y'$ in the massless limit, where $Y'$ is a compact Calabi-Yau 3-fold.
This superspace was used in \cite{Buchbinder:2003qu} to compute supergravity loop corrections to supersymmetry breaking in (a phenomenological analog of) heterotic M-theory on $Y'$.}

Alternatively, one may attempt to define the theory in a gravitino superfield $\Psi(x,y,\theta,\bar \theta)$ expansion keeping only the 4D, $N=1$ supergravity fields and the superfields holding the components of the 3-form but all non-linearly.
In such an approach, we expect the action to take the form 
\begin{align}
\label{E:GravitinoExpansion}
S = S_{CS} +
	S_K
	+\mathcal O(\Psi) 
\end{align}
to lowest order.
Here, the Chern-Simons action $S_{CS}$ is taken to be the invariant of the non-abelian tensor hierarchy constructed in \cite{Becker:2016xgv,Becker:2016rku}. 
(This hierarchy encodes the components of the dimensionally-decomposed eleven-dimensional 3-form.)
We will refer to the remaining terms $S_K$ as the ``K\"ahler action''. 
We propose to take it to be a natural generalization of the super-volume on $\bm X\times Y$ constructed from the remaining supergravity and tensor hierarchy fields (cf.\ eq.\ \ref{E:Kahler}).

Additionally, one may consider freezing the 4D, $N=1$ supergravity multiplet around a flat $\mathbf R^{4|4}$ background and letting only the tensor hierarchy fields fluctuate. (We take the spacetime part to be flat for simplicity, but a curved rigid background may be considered instead.) In this approximation, the K\"ahler term reduces to the superfield analog of the Riemannian volume on $Y$.
In our approach, the metric scalars are the imaginary part of chiral scalar fields in the tensor hierarchy that are 3-forms on $Y$. (The real part holds the 3-form scalars.) This defines a $G_2$ structure on $Y$.
The result of this is that the K\"ahler action is, essentially, a superspace lift of the Hitchin functional \cite{Hitchin:2000jd, Hitchin2001}. 

In this work, we test this proposal by computing the scalar potential of this action. 
As we are freezing the spactime supergravity part, this background will be of the form $\mathbf R^{4|4} \times Y$ with $Y$ a $G_2$ structure manifold (not necessarily compact).
The remaining fields are those of the non-abelian tensor hierarchy. 
In particular, we reproduce the scalar potential of eleven-dimensional supergravity from the Chern-Simons action and the Hitchin functional.
This potential consists of the Ricci scalar on $Y$ in a form due to Bryant \cite{Bryant2003} and an analogous expression for the square of the Maxwell-like tensor for the 3-form scalars. 

Before concluding, let us pause to compare the proposed set-up to the analogous construction for ten-dimensional, $N=1$ super-Yang-Mills worked out by Marcus, Sagnotti, and Siegel \cite{Marcus:1983wb}.
In that case, the superspace is of the form $\mathbf R^{4|4}\times Y'$ where $Y'$ has a fixed $SU(3)$ structure $(\bar \partial , \Omega)$. 
The components of the ten-dimensional gauge field are embedded in a real superfield (1-form components along $X$) and three chiral superfields $\phi_i$ (components along $Y'$) transforming in the $\mathbf 3$ of $SU(3)$. All four superfields are valued in the adjoint representation of the gauge group $G$. 
The superpotential of the theory is the superspace generalization of holomorphic Chern-Simons functional
\begin{align}
W_{MSS} = \frac12 \int d^4x \int d^2 \theta  \int_Y   \bar \Omega \wedge
	\mathrm{Tr} \, \left(
		\phi \wedge \partial\phi + \frac {2}3 \phi \wedge \phi \wedge \phi
	\right)
\end{align}
with the trace taken in the adjoint representation of $G$.
The F-term condition that follows from this action implies the vanishing of the $(2,0)$ part of the Yang-Mills field strength so that (the lowest component of) $\phi$ describes an anti-holomorphic connection.
In this analogy, the vector multiplet we are ignoring plays the role of the gravitational fields we are freezing (with indices along $X$) and the chiral scalars $\phi$ stand in for the scalar fields in the tensor hierarchy (all indices along $Y$).

In the next section, we describe the embedding of the components of eleven-dimensional supergravity into superfields on $\mathbf R^{4|4} \times \mathbf R^7$. (The fields on $\bm X\times Y$ follow from this by covariantizing derivatives, as usual \cite{Gates:1983nr, Wess:1992cp, Buchbinder:1998qv}.)
In section \ref{S:Action} we construct the action from the Chern-Simons invariant of the non-abelian tensor hierarchy and a supersymmetric extension of the Hitchin functional for $Y$. 
The equations of motion of the F- and D-auxiliary fields are computed. 
From this we obtain simultaneously the conditions for supersymmetry and the scalar potential. 
We conclude in section \ref{S:Discussion} with a discussion of our results. 
Appendix \ref{S:G2} contains a brief review of $G_2$ structures, the Hitchin functional, and some useful identities.

\section{Superfields and Components}
\label{S:Fields}
We begin by embedding the eleven-dimensional component fields into simple superspace. 
The eleven-dimensional supergravity component spectrum consists of a Riemannian metric $g_{\bm {mn}}$ or, more properly, its frame $e_{\bm m}{}^{\bm a}$, a 32-component Majorana gravitino $\psi_{\bm m}^{\bm \alpha}$, and an abelian 3-form gauge field $C_{\bm{mnp}}$. 
Here, bold indices are eleven-dimensional: ${}_{\bm \alpha},{}_{\bm \beta} = {}_{1,\dots, 32}$ are Majorana spinor indices and we use the early-late convention for tangent vector indices ${}_{\bm a},{}_{\bm b} = {}_{0,\dots, 9}$ and coordinate indices ${}_{\bm m},{}_{\bm n} = {}_{0,\dots, 9}$. 
The bosonic part of the eleven-dimensional supergravity action is given by
\begin{align}
\label{E:11DSG}
\kappa^ 2 S_{11} = \int d^{11}x \sqrt{-g} \left[
	 \frac12  R (g)
	-\frac1{4\cdot 4!} F_{\bm{mnpq}}^2 	
	\right]
	-\frac1{12} \int C \wedge F \wedge F 
.	
\end{align}
Here, $R$ is the Ricci scalar of the metric, $g$ is its determinant, and $F=dC$ is the 4-form field strength of the gauge 3-form $C$.

As we will be embedding into a superspace modeled on $\mathbf R^{4|4} \times \mathbf R^7$, we must first reduce these components to ``spacetime'' $X$ and ``internal'' $Y$:
\begin{align}
\label{E:components}
e_m{}^a
~,~~
e_m{}^i
~,~~
g_{ij}
~,~~
\psi_m^{\alpha I}
~,~~
\psi_i^{\alpha I}
~,~~
C_{mnp} 
~,~~
C_{mn\, i}
~,~~
C_{m\, ij}
~,~~
C_{ijk}
~.~~
\end{align}
The new indices on $X\times Y$ are as follows: ${}_m,{}_n = {}_{0,\dots, 3}$ denote spacetime coordinate indices, ${}_a,{}_b = {}_{0,\dots, 3}$ are spacetime tangent vectors indices, ${}_i,{}_j = {}_{1,\dots, 7}$ will be taken to be internal coordinate indices, ${}_{\alpha} , {}_{\dt \alpha} = {}_{1,2}$ are $SL(2, \mathbf C)$ indices, and finally, ${}^I{}^{\textrm,}{}^J = {}^{1,\dots, 8}$ stand for $SO(8)$ R-symmetry indices. 

To embed in superfields of $\mathbf R^{4|4} \times \mathbf R^7$, it is necessary to split up the gravitino fields and put one of them into an irreducible superspin-$\frac32$ multiplet with the frame $e_m{}^a$. This will then be the 4D, $N=1$ super-frame $E_M{}^A$. The other 7 gravitini must then go into a superspin-1 multiplet $\Psi^{\alpha i}$ transforming in the defining representation of the $SO(7)\subset SO(8)_R$ subgroup. (For notational simplicity, we do not distinguish between coordinate indices on $Y$ and this subgroup.) 
The remaining fields consist of 1 3-form, 7 2-forms, $21+7= 28$ vectors, and $28+35=63$ scalars (and their spin-$\frac12$ superparters). 
This set of fields is encoded in a non-abelian tensor hierarchy \cite{Becker:2016xgv, Becker:2016rku} as we review presently.

Since there are many (super)fields involved, we try to minimize notation as follows:
For any superfield $X$ we define supersymmetry-covariant descendant superfields by acting with superspace derivatives. The descendants with the same statistics as $X$ are defined by
\begin{align}
\label{E:ComponentsSame}
f_X = -\tfrac14 \bar D^2 X
~&,~~
A_{Xa} = -\tfrac14 (\tilde \sigma_a)^{\dt \alpha \alpha} [D_\alpha, \bar D_{\dt \alpha}] X
~,~~
\cr
\tilde f_X = -\tfrac14 D^2 X 
~&,~~
d_X = \tfrac1{32} \{D^2, \bar D^2\} X ~,
\end{align}
whereas those of opposite statistics are
\begin{align}
\label{E:ComponentsOpposite}
\chi_{X\alpha} = D_\alpha X
~&,~~
w_{X \alpha } = -\tfrac14 \bar D^2 D_\alpha X
~,~~
\cr
\tilde \chi_{X\dt \alpha} = \bar D_{\dt \alpha} X
~&,~~
\tilde w_{X \dt \alpha } = -\tfrac14 D^2 \bar  D_{\dt \alpha} X 
~.~~
\end{align}
These superfields are used to define the covariant component fields by projecting $\theta, \bar \theta \to 0$ an operation we denote with a ``$|$''.
In terms of these components, the superfield can be written as
\begin{align}
X=X|
	&+\theta^{\alpha}\chi_{X_{\alpha}}| 
	+ \bar{\theta}_{\dt{\alpha}}\tilde{\chi}^{\dt{\alpha}}_{X}|
	+\theta^2f_{X}|
	+\bar{\theta}^2\tilde{f}_{X}| 
	- \theta \sigma^a \bar \theta A_{Xa}|
\\	&+ \bar{\theta}^2\theta^{\alpha}
		\left(
		w_{X \alpha}|+\tfrac{i}{2}\sigma^a_{\alpha \dt{\alpha}}\partial_a\tilde{\chi}_X^{\dt{\alpha}}|
		\right)  
	+\theta^2\bar{\theta}_{\dt{\alpha}}(\tilde{w}^{\dt{\alpha}}_X|
	+\frac{i}{2}\tilde{\sigma}^{a \dt{\alpha}\alpha }\partial_a\chi_{X_{\alpha}})|
	+\theta^2\bar{\theta}^2\left({d_X}|-\tfrac{1}{4}\Box X|\right)
.
\nonumber
\end{align}
Henceforth, we will drop the ``$|$'' notation on the right-hand side of such expansions.
When $X$ is real, the tilded fields are conjugate to the untilded ones and $X$, $A_X$, and $d_X$ are real. When $X$ is chiral, the tilded fields are absent and the remaining components are complex. 

In this work, we will not have much need for the superspin-$s$ fields with $s=\tfrac32$ (superframe $E_M{}^A$) and 1 (seven gravitino superfields $\Psi^{\alpha i}$) so we will be brief. 
(For an explicit construction of the quadratic action of 5D, $N=1$ supergravity analog in terms of these superfields, see refs.\ \cite{Linch:2002wg, Gates:2003qi}.)
At the linearized level, the conformal graviton can be described by the real superfield 
\begin{align}
\label{E:H}
H^a &= \dots + \theta \sigma^m \bar \theta e_m{}^a 
	+ \bar \theta^2 (\sigma^{ab}\theta)_\alpha \psi_b^\alpha
	+ \theta^2 (\tilde \sigma^{ab}\bar \theta)_{\dt \alpha} \bar \psi_b^{\dt \alpha}
	+ \theta^2 \bar \theta^2 d^a
.	
\end{align}
It contains the (linearized) frame, the $N=1$ gravitino, and a real auxiliary vector field. 
(Here and in the following ellipses will stand for components that can be removed by a choice of Wess-Zumino gauge.)
The gravitino superfield 
\begin{align}
\label{E:Gravitino}
\Psi^{\alpha i} &= \dots 
	+ (\sigma^a\bar \theta)^\alpha B_a{}^i
	+(\theta \sigma^m\bar \theta) \psi_m^{\alpha i}
	+ \bar \theta^2 \theta^\alpha w^i
	+ \theta^2 (\sigma^a\bar \theta)^\alpha y^i_a
	+ \bar \theta^2 (\theta \sigma^{ab})^\alpha w^i_{ab}
	+ \dots 
\end{align}
carries the 7 remaining gravitini, 7 ``graviphotons'', and a collection of auxiliary fields, the precise content of which depends on the structure of the supergravity gauge transformation (which we will not need). 

All but one of the remaining bosonic fields can be embedded in an abelian tensor hierarchy of superfields  \cite{Becker:2016xgv}.
This is a chain complex of superfields constructed by combining the superspace analog of the de Rham complex on $X$ and the de Rham complex on $Y$. The components of the 11D 3-form $C_{\bm{abc}}$ fit into the elements of this complex as follows:
\begin{subequations}
\label{E:ATHprepotentials}
\begin{align}
\Phi_{ijk} &= C_{ijk} + i {F}_{ijk}  + \dots + \theta^2 f_{ijk}
\\
V_{ij} &= \dots + \theta \sigma^a \bar \theta C_{a\, ij} + \dots + \theta^2 \bar \theta^2 d_{ij}
\\
\label{E:Sigma}
\Sigma_{\alpha i} &= \dots + \theta_\alpha H_i + (\theta \sigma^{ab})_\alpha C_{ab\, i} + \dots
\\
\label{E:X}
X &= \dots+ \bar\theta^2  G+ \theta^2 \bar G + \theta \sigma_a \bar \theta \epsilon^{abcd} C_{bcd}  + \dots + \theta^2 \bar \theta^2 d_X
.
\end{align}
\end{subequations}
General abelian Chern-Simons-like invariants of this hierarchy in superspace were constructed in \cite{Becker:2016xgv}.
For eleven-dimensional supergravity, this is a cubic invariant of this superspace complex.

Being valued in the exterior algebra of the internal space, these fields are ``charged'' under the mixed components of the frame gauging the $\mathfrak g=\mathfrak{diff}(Y)$ symmetry. This results in a non-abelian gauging of the tensor hierarchy by a super-$\mathfrak g$-connection with spinorial superfield
\begin{align}
\label{E:NonAb}
\mathcal A_\alpha^i &= \dots + (\sigma^a\bar \theta)_\alpha e_a{}^i
	+\dots
	+ \theta_\alpha \bar \theta^2 \bm d^i
\end{align}
arising by gauge-covariantizing the flat superspace derivative $D_\alpha \to \mathcal D_\alpha$ (minimal coupling). 
At this point the non-abelian tensor hierarchy is a $\mathfrak g$-equivariant super-de Rham complex of forms on $\bm X \times Y$. Its Chern-Simons-like invariant was studied in some detail and generality in \cite{Becker:2016rku}. 
We review the gauge transformations, $\mathfrak g$-covariant superfield strengths, Bianchi identities, and Chern-Simons action in section \ref{S:CSAction}.

It remains to discuss the fate of the 28 metric scalars $g_{ij}$. 
Although we have not embedded them explicitly, we expect that they can be accounted for by the real scalars $F_{ijk}$ in the chiral field of the tensor hierarchy. 
(We elaborate on this in section \ref{S:KahlerAction}.)
Assuming this, we have embedded the component fields (\ref{E:components}) of eleven-dimensional supergravity into a collection of prepotentials consisting of the conformal supergraviton $H^a$ (\ref{E:H}), 7 gravitino superfields $\Psi^{\alpha i}$ (\ref{E:Gravitino}), and the superfields of the gauged tensor hierarchy (\ref{E:ATHprepotentials} and \ref{E:NonAb}).
As we will be freezing the conformal graviton and gravitino superfields, the remaining set of auxiliary fields come only from the tensor hierarchy. They consist of the components
\begin{align}
\label{E:Auxiliary}
d_X
~,~~ 
\bm d^i = d_{\mathcal V^i}
~,~~ 
d_{ij} &= d_{V_{ij}}
~,~~
f_{ijk}=f_{\Phi_{ijk}}
.
\end{align}
(Here $\mathcal V^i$ is the prepotential of the non-abelian gauge field $\mathcal A^i_\alpha \partial_i \sim e^{- i \mathcal V\partial}(D_\alpha e^{ i \mathcal V\partial} )$.) 
In the next section, we will propose an action constructed from the superfields of this section and project to components, focusing on this set of auxiliary fields. 

\section{Action}
\label{S:Action}

In this section we propose an action constructed from the supergravity and tensor hierarchy fields to lowest order in a gravitino superfield expansion (\ref{E:GravitinoExpansion}). This is a superspace action consisting of a Chern-Simons term and a generalization of the Hitchin functional. 

\subsection{Chern-Simons Action}
\label{S:CSAction}
The superspace Chern-Simons action appropriate to eleven-dimensional supergravity is the cubic invariant of the non-abelian tensor hierarchy describing a gauge 3-form. 
The embedding of the components of this 3-form into superfields is represented in equation (\ref{E:ATHprepotentials}).
The gauge transformations for these prepotentials are\footnote{We attempt to give a self-contained description of that part of the work on Chern-Simons-like invariants of the non-abelian tensor hierarchy needed for this paper, but please see references \cite{Becker:2016xgv,Becker:2016rku} for additional material and information on this important ingredient of the construction. 
} 
\begin{subequations}
\label{E:NATHtransformations}
\begin{align}
\delta \Phi &= 
	\mathscr L_\lambda \Phi 
	+ \partial \Lambda 
\\
\delta V &= 
	\mathscr L_\lambda V 
	+ \tfrac1{2i}\left(\Lambda - \bar \Lambda \right)
	- \partial U 
\\
\delta \Sigma_\alpha &= 
	\mathscr L_\lambda \Sigma_\alpha 
	-\tfrac14 \bar {\mathcal D}^2 \mathcal D_\alpha U 
	+ \partial \Upsilon_\alpha 
	+ \iota_{\mathcal W_\alpha} \Lambda
\\
\delta X &= 
	\mathscr L_\lambda X
	+\tfrac1{2i}\left(\mathcal D^\alpha \Upsilon_\alpha - \bar {\mathcal D}_{\dt \alpha} \bar \Upsilon^{\dt \alpha} \right)
	- \omega_{\mathsf h}(\mathcal W_\alpha, U) 
.
\end{align}
\end{subequations}
All fields are differential forms on $Y$; $\partial$ denotes the de Rham differential and wedge products are implied. 
The abelian part of the gauge transformation is parameterized by the superfields $\Lambda_{ij}$ (chiral), $U_i$ (real), and $\Upsilon_\alpha$ (chiral) encoding the components of an eleven-dimensional super-2-form.
The non-abelian part $\mathfrak g = \mathfrak {diff}(Y)$ acts by the Lie derivative with respect to the real scalar superfield $\lambda^i$.
To check gauge invariance, we must use separation of the Lie derivative into the de Rham differential and the contraction operator $\iota$ using Cartan's formula $\mathscr L_\mathcal V = \partial \iota_\mathcal V + \iota_\mathcal V \partial$. 
The composite superfield $\omega_{\mathsf h}$ is the so-called ``Chern-Simons superform''. For any chiral spinor superfield $\chi_\alpha$ and real scalar superfield $v$,
\begin{align}
\omega_{\mathsf h}(\chi_\alpha, v) &:= 
\iota_{\chi^\alpha} \mathcal D_\alpha v 
	+ \iota_{\bar \chi_{\dt \alpha}} \bar {\mathcal D}^{\dt \alpha} v 
	+\tfrac12 \left(  \iota_{\mathcal D^\alpha\chi_\alpha} v 
		+ \iota_{\bar {\mathcal D}_{\dt \alpha} \bar \chi^{\dt \alpha}} v 
	\right)
\cr
~~~\Rightarrow~~~
\bar {\mathcal D}^2 \omega_{\mathsf h}(\chi_\alpha, v) &= 
\iota_{\chi^\alpha} \bar {\mathcal D}^2 \mathcal D_\alpha v 
	+ \tfrac12 \bar {\mathcal D}^2  \iota_{\mathcal D^\alpha \chi_\alpha - \bar {\mathcal D}_{\dt \alpha} \bar \chi^{\dt \alpha}} v 
.
\end{align}
(Its name derives from the fact that if $\chi \sim \bar D^2 D v$ is the field strength superfield of the real vector superfield $v$, then the second term vanishes and $\bar D^2 \omega \sim \chi^2$ gives the superspace analog of $d \omega = F \wedge F$.)

The non-abelian gauge field strength $\mathcal W^{\alpha i}$ is defined by $[\mathcal D_a, \bar {\mathcal D}_{\dt \alpha}] = (\sigma_a)_{\alpha \dt \alpha} \mathcal L_{\mathcal W^{\alpha}}$. 
The field strengths $\partial \Phi$ and 
\begin{subequations}
\label{E:NATHFS}
\begin{align}
\label{E:FSF}
F &= \tfrac1{2i}\left( \Phi - \bar \Phi\right) - \partial V
\\
W_\alpha &= -\tfrac14 \bar {\mathcal D}^2 \mathcal D_\alpha V 
	+\partial \Sigma_\alpha 
	+ \iota_{\mathcal W_\alpha} \Phi
\\
H &= \tfrac1{2i}\left(\mathcal D^\alpha \Sigma_\alpha - \bar {\mathcal D}_{\dt \alpha} \bar \Sigma^{\dt \alpha} \right)
	-\partial X 
	-\omega_{\mathsf h}(\mathcal W_\alpha, V)
\\
\label{E:G}
G&= -\tfrac14 \bar {\mathcal D}^2 X 
	+ \iota_{\mathcal W^\alpha} \Sigma_\alpha
\end{align}
\end{subequations}
are invariant under the abelian transformations and covariant under the non-abelian ones: $\delta (FS) = \mathscr L_\lambda (FS)$.
Being given explicitly in terms of the prepotential superfields, these field strengths identically satisfy the Bianchi identities
\begin{subequations}
\label{E:NATHBI1}
\begin{align}
\tfrac1{2i}\left(\partial \Phi - \partial \bar \Phi\right) &=  \partial F
	\\
-\tfrac14 \bar {\mathcal D}^2 {\mathcal D}_\alpha F  &= 
	-\partial W_\alpha 
	- \iota_{\mathcal W_\alpha} \partial \Phi
	\\
\label{E:NATHBI1W}
\tfrac1{2i}\left({\mathcal D}^\alpha W_\alpha - \bar {\mathcal D}_{\dt \alpha} \bar W^{\dt \alpha} \right)&= 
	\partial H 
	+ \omega_{\mathsf h}(\mathcal W_\alpha, F)
	\\ 
-\tfrac14 \bar {\mathcal D}^2 H &= 
	-\partial G 
	- \iota_{\mathcal W^\alpha} W_\alpha 
	\\
\bar {\mathcal D}_{\dt \alpha} G &= 0.
\end{align}
\end{subequations}
(Equivalently, \ref{E:NATHFS} is the solution to these constraints.)
In terms of these prepotentials and field strength superfields, the Chern-Simons super-invariant $S_{CS} = \int d^4 x \int_Y L_{CS}$ is defined by the Lagrangian
\begin{align}
\label{E:NACS}
-12\kappa^ 2 L_{CS} &= 
	i \int d^2 \theta \, \Phi \wedge \left[ 
		 \partial \Phi G
	+\tfrac 12 W^\alpha \wedge W_\alpha
	-\tfrac i4 \bar {\mathcal D}^2\left( F \wedge  H \right)
	\right]
\cr		
	&
	+ \int d^4 \theta \, V \wedge \left[ 
		\partial \Phi \wedge H
		+  F\wedge \mathcal D^\alpha W_\alpha
		+ 2\mathcal D^\alpha F\wedge \left(W_\alpha 
		-i \iota_{\mathcal W_\alpha} F \right)
		\right]
\cr		
	&
	+ i \int d^2 \theta \, \Sigma^\alpha \wedge \left[ 
	  	 \partial \Phi\wedge  W_\alpha 
	-\tfrac i4 \bar{\mathcal D}^2 \left( F\wedge  {\mathcal D}_\alpha F\right)
	\right]
\cr
	&
	-  \int d^4 \theta \, X \partial \Phi\wedge  F
+\mathrm{h.c.} 	
\end{align}
Using the Bianchi identities, one can check that this Lagrangian transforms into an exact superform under the gauge transformations (\ref{E:NATHtransformations}). 
Comparing the first term with the superpotential $\int d^2 \theta \Phi \partial \Phi$ put forward in reference \cite{Becker:2014rea} suggests an interpretation of this action as the covariantization of the $G_2$ superpotential under the tensor hierarchy transformations.

The tensor hierarchy can be coupled to gravity by replacing $\mathcal D\to \mathscr D$ with the gravitationally covariant superspace derivative, covariantizing the measures as usual, and replacing $\bar {\mathcal D}^2 \to \bar{\mathscr D}^2 - 8 R$ \cite{Randall:2016zpo, Randall:Note}. (Equivalently, one can replace $\mathcal D$ by the conformal superspace supergravity derivative \cite{Aoki:2016rfz, Yokokura:2016xcf}.)
Then, the complete component projection of the action can be computed straightforwardly, but for this paper we are interested only in the contribution to the scalar potential of the component theory. 
This simplifies the calculations significantly. 

Firstly, we can ignore the supergravity couplings so the action in the form (\ref{E:NACS}) will suffice to compute the potential. 
Next, the potential consists of only internal derivatives of the scalars $g_{ij}$ and $C_{ijk}$ in the hierarchy so we will drop all gauge fields (with $X$ indices) and spacetime derivatives. (Of course there are internal derivatives of gauge fields but we find it more convenient to think of them as covariantizing the spacetime derivatives.)
The gauge superfields can still contribute F- and D-type auxiliary fields so we will keep those. 
An exception is the superfield $\Sigma_{\alpha i}$ for the gauge 2-form $C_{ab i}$: This representation (\ref{E:Sigma}) has no auxiliary fields so we will remove it altogether. (We revisit the validity of this simplification in section \ref{S:Discussion}.)
Performing the Gra\ss{}mann integration, and focusing only on the remaining fields, we find 
\begin{align}
\label{E:CScomponents1}
\kappa^2 L_{CS} = \tfrac1{288} \epsilon^{ijklmnp} \Big[
 F_{ijkl} {F}_{mnp} d_X
&-4{F}_{ijk} {\bm d}^r {F}_{lm r} {\bm d}^s{F}_{np s}
+12{F}_{ijk} d_{lm}d_{np} 
\cr
	&-\tfrac12 \Big(G[4{F}_{ijk}\partial_l f_{mnp} 
	+ i F_{ijkl}f_{mnp}]
	+\mathrm{h.c.}\Big) 
	\Big] .
\end{align}
Here $F_{ijkl}= 4\partial_{[i}C_{jkl]}$ is the lowest component of the real part of $\partial \Phi$.\footnote{This should not be confused with $F_{ijk}$ which denotes the lowest component of the field strength (\ref{E:FSF}). ($G$ is the lowest component of (\ref{E:G}), and the remaining auxiliary fields are defined in (\ref{E:Auxiliary}).)}
We use this result in section \ref{S:ScalarPotential} once we have constructed the remaining terms to which we turn next.

\subsection{K\"ahler Action}
\label{S:KahlerAction}
In the previous section, we reviewed the Chern-Simons action arising in the gravitino superfield expansion 
(\ref{E:GravitinoExpansion}). 
The remaining terms at this order in the expansion define the K\"ahler action $S_K$. 
In this section, we propose an explicit formula for this action. 

In the component embedding of section \ref{S:Fields}, we assumed the metric scalars (pull-back of the 11D metric to $Y$) appeared in the superfield spectrum in the imaginary part of the 35 chiral fields carrying the 3-form scalars (the real part).
In section \ref{S:CSAction}, we reviewed the construction of the superfield strength $F_{ijk}$ associated to these chiral fields (eq.\ \ref{E:FSF}).
This field strength is invariant under the abelian 3-form transformations as should be the metric scalars. 
Therefore, we identify the dynamical metric scalars as the fluctuations of this field strength and construct a Riemannian metric following Hitchin: First, define the superfield
\begin{align}
s_{ij}(F) := -\frac1{144} \epsilon^{klmnpqr}F_{ikl} F_{jmn} F_{pqr} 
.
\end{align}
Stability of $F$ means that $\mathrm{det}(s)\neq 0$. 
Setting $g_{ij}(F) := \mathrm{det}^{-1/9}(s) s_{ij}$, defines a Riemannian metric on $Y$.
We note that the replacement $\Phi_{ijk}\to F_{ijk}$ guarantees abelian gauge invariance but it is not holomorphic.
This has the important consequence that we cannot use $g_{ij}(F)$ to construct invariant superpotential terms.

Taking the determinant, we can define the superspace analog of the Riemannian volume on $Y$. 
So motivated, we propose to take as the K\"ahler action the following natural generalization of the
superspace volume of $\bm X\times Y$:
\begin{align}
\label{E:Kahler}
S_K = -\frac3{\kappa^2} \int d^4 x \int d^7y \int d^4\theta \,E  \, \sqrt{g(F)} \, \Big[
	(\bar GG)^{1/3}
	-\tfrac13 (\partial_i H_a)^2
	\Big]
.	
\end{align}
Here, $E = \mathrm {sdet}(E_M{}^A)$ is the super-determinant of the 4D, $N=1$ components of the frame.
The second term is the $N=1$ super-graviton ``mass'' term. 
(This term is needed because there are no spin-2 auxiliary fields.)
Its normalization is fixed by eleven-dimensional Lorentz invariance: 
In the quadratic approximation,
$- 3 \int d^4 x d^7y \, d^4\theta \, E \to  - \int d^4 x d^7y \,  H_a \Box H^a$ in the gauge $\bar D_{\dt \alpha} H^{\un a} = 0$ \cite{Linch:2002wg, Gates:1983nr, Buchbinder:1998qv}.

The first term looks like the old-minimal supergravity action \cite{Gates:1983nr, Buchbinder:1998qv}, provided we identify $G\leftrightarrow \Phi_0$ with the chiral conformal compensator.\footnote{Actually, this would be a slightly modified version of old-minimal supergravity since $G$ has a real prepotential $X$ (\ref{E:G} and \ref{E:X}). Such a modification of the compensator was first exploited in \cite{Grisaru:1981xm} to simplify supergraph calculations. It was also needed in the (4+1)-dimensional version of the superspace supergravity considered here \cite{Linch:2002wg, Buchbinder:2003qu, Gates:2003qi}.
} 
We will revisit this connection in section \ref{S:Discussion} but if we simply assume it for now, then
freezing gravity amounts to setting 
\begin{align}
\label{E:Freezing}
H^a \to \theta \sigma^a \bar \theta 
~~~,~~~
\Psi^{\alpha i} \to 0
~~,~\mathrm{and}~~~ 
G\to 1+ \theta^2 (d_X - \tfrac i{4!} \epsilon^{abcd} F_{abcd}).
\end{align}
As we are interested only in the scalar potential in this paper, we drop the field strength of the component 3-form $C_{abc}$. (However, matching the coefficient of this component to the correct value fixes the $G$-dependence in eq.\ \ref{E:Kahler}.
In the setting in which 4-form fluxes are turned on, these terms give quantum corrections to the potential as explained by Beasley and Witten \cite{Beasley:2002db}.)
This reduces the K\"ahler part of the action to
\begin{align}
\label{E:Hitchin}
S_K\to  - \frac 3{\kappa^2} \int d^4 x \int d^7y \int d^4\theta \, \sqrt{g(F)} \, 
	\left[ 1 +\tfrac13(\theta^2 + \bar \theta^2) d_X + \tfrac19\theta^2 \bar \theta^2 d_X^2 
	\right]
.	
\end{align}
We recognize the leading term as a superspace version of the Hitchin functional (\ref{E:HitchinFunctional}) for the $G_2$ structure on $Y$ \cite{Hitchin:2000jd, Hitchin2001} (see also \cite{Dijkgraaf:2004te}).

Integrating over the odd coordinates and collecting the auxiliary field terms needed to compute the scalar potential, we find
\begin{align}
\label{E:KahlerComponents1}
\kappa^2 L_K = 
	-\tfrac13 \sqrt{g} d_X^2 
	-\tfrac 1{18} \sqrt{g} {F}^{ijk} \mathrm{Im}(f_{ijk})d_X
	&+ \tfrac16 \sqrt{g} {F}^{ijk} F_{ijkl}\bm d^l
\\	&-\tfrac12 \sqrt{g} {F}^{ijk} \partial_k d_{ij}
	+\tfrac 34 G^{ijk, mnp} \bar f_{ijk}f_{mnp}
.
\nonumber
\end{align}
Here, 
\begin{align}
\label{E:HitchinMetric}
G^{ijk,mnp} := 
	 	\tfrac1{3!\cdot 3! } \sqrt{g} g^{[i|m}g^{|j|n}g^{|k]p}
	 	+\tfrac1{18\cdot 3!\cdot 3! }\sqrt{g} {F}^{ijk}{F}^{mnp}
		+\tfrac1{4!} \sqrt{g} g^{[m|[i}\psi^{jk]|np]}
\end{align}
is essentially the Hitchin metric on the moduli space of (complexified) $G_2$ structures \cite{Hitchin:2000jd}.
In terms of $G_2$ projections
\begin{align}
\label{E:H3form}
18 \omega_{ijk} G^{ijk,lmn} \omega_{lmn} &= 
	-\tfrac43 \omega_{\mathbf 1 ijk}^2
	-\omega_{\mathbf 7 ijk}^2
	+\omega_{\mathbf{27} ijk}^2
\end{align}
for any 3-form $\omega$.

\section{Scalar Potential}
\label{S:ScalarPotential}

Using various identities from $G_2$ linear algebra collected in appendix \ref{S:G2}, we can rewrite the Chern-Simons contribution (\ref{E:CScomponents1}) as
\begin{align}
\label{E:CScomponents2}
\kappa^2 L_{CS} &=  2\sqrt{g} g_{ij} \bm d^i \bm d^j
	+\tfrac14\sqrt{g}\psi^{ijkl} d_{ij} d_{kl}	 
+\tfrac1{48}\sqrt{g}\psi^{ijkl}F_{ijkl}d_X
\\ 
&-\tfrac1{12}\partial_i(\sqrt{g} \psi^{ijkl}) \mathrm{Re}(f_{jkl})
	+\tfrac 1{288} \epsilon^{ijklmnp}
	F_{ijkl}\mathrm{Im}(f_{mnp})
	.
\nonumber
\end{align}
To proceed, it is useful to introduce the intrinsic torsion forms $\tau_\mu$ for $\mu=0,1,2,3$ and analogous quantities $\sigma_\mu$ for $\mu=0,1,3$ defined by \cite{Bryant2003}
\begin{subequations}
\begin{align}
\label{E:TorsionClassesDef}
d\varphi &= \tau_0 \psi + 3\tau_1\varphi + \ast \tau_3
~,~~
d\psi = 4 \tau_1 \psi + \tau_2 \varphi 
~,
\\
dC &= \sigma_0 \psi + 3\sigma_1\varphi + \ast \sigma_3
~.
\end{align}
\end{subequations}
(We could make the analogous definition for the components of $d \ast C$ but the action depends only on $C$ and $dC$; the $C$-field analogue of the torsion class $\tau_2$ is not gauge invariant.)

In terms of these intrinsic torsion forms, $\tfrac1{\sqrt{g}}  \partial_k \left( \sqrt{g} {F}^{ijk} \right) = 4 {F}^{ijk} (\tau_1)_k + (\tau_2)^{ij}$
which we use to rewrite (\ref{E:KahlerComponents1}) as
\begin{align}
\label{E:KahlerComponents2}
\kappa^2 L_K= 
	-\tfrac13 \sqrt{g} d_X^2 
	-\tfrac 1{18} \sqrt{g} {F}^{ijk} \mathrm{Im}(f_{ijk})d_X
	- 2\sqrt{g} d^{ij} \left [{F}_{ijk}(\tau_1)^k
	+ \tfrac14 (\tau_2)_{ij} \right]
\cr
	+ 12\sqrt{g}\bm d^i (\sigma_1)_i
	+\tfrac34\sqrt{g} \bar f_{ijk} G^{ijk , mnp} f_{mnp}
.	
\end{align}
In terms of $G_2$ representations this is becomes
\begin{align}
\label{E:KahlerComponents3}
\kappa^2 L_K= 
	-\tfrac13 \sqrt{g} d_X^2 
	-\tfrac 1{18} \sqrt{g} {F}^{ijk} \mathrm{Im}(f_{\mathbf 1 ijk})d_X
	- 2\sqrt{g} d^{ij} \left [{F}_{ijk}(\tau_1)^k
	+ \tfrac14 (\tau_2)_{ij} \right]
\cr
	+ 12\sqrt{g}\bm d^i (\sigma_1)_i
	-\tfrac1{18}\sqrt{g} |f_{\mathbf 1 ijk}|^2
		-\tfrac1{24}\sqrt{g} |f_{\mathbf 7 ijk}|^2
		+\tfrac1{24}\sqrt{g} |f_{\mathbf {27} ijk}|^2
.
\end{align}
Combining this result with (\ref{E:CScomponents2}), we find for the equations of motion of the auxiliary fields
\begin{subequations}
\begin{align}
d_{X} &= \tfrac{21}4 \sigma_0 	
	-\tfrac 1{12} {F}^{ijk} \mathrm{Im}(f_{\mathbf 1 ijk})
\\
g_{ij} \bm d^j  &= 
	 - 3(\sigma_1)_{i}
\\
\psi_{ijkl} d^{kl}  &= 
	  4{F}_{ijk}g^{kl}(\tau_1)_{l}
	+ (\tau_2)_{ij} 
\\
G^{ijk ,lmn} \bar f_{lmn} &= 	 
	\tfrac i{432}\epsilon^{ijklmnp} F_{lmnp}
	-\tfrac1{18} \partial_l\psi^{ijkl}
	-\tfrac i{27}  d_X F^{ijk}
\end{align}
\end{subequations}
To solve these, it is convenient to decompose into $G_2$ representations and invert on irreducible representations using the identities in appendix \ref{S:G2}. 
Doing so gives
\begin{align}
\label{E:AuxiliaryEOM}
d_{X} = - \tfrac72 \sigma_0 
~,~~  
	\bm d^i = - 3(\sigma_1)^i 
~&,~~
d_{\mathbf 7 ij} = -{F}_{ijk}(\tau_1)^k
~,~~  
	d_{\mathbf {14} ij} =  \tfrac12 (\tau_2)_{ij} 
~,~~\\
\bar f_{\mathbf{1}ijk} = 
	-\left( \tfrac34\tau_0 + \tfrac52 i\sigma_0 \right)F_{ijk} 
~,~~
\bar f_{\mathbf{7}ijk} &= 
	 3 \psi_{ijkl} (\tau_1 +i \sigma_1)^l 
~,~~  
\bar f_{\mathbf{27}ijk} = 
	(\tau_3 +i \sigma_3)_{ijk}
~.~	
\nonumber
\end{align}
Note that the F- and D-flatness conditions are equivalent to the vanishing of each of the intrinsic torsion forms ($\tau_\mu=0$) and their gauge field counterparts ($\sigma_\mu=0$ for $\mu\neq 2$).

In terms of $G_2$ projections, we may write the potential in terms of auxiliary fields as
\begin{align}
\label{E:ScalarPotential1}
V&=
	\tfrac14d_X^2 
	+ 2 g_{ij} \bm d^i \bm d^j 		
	-  d_{\mathbf 7 ij} ^2 
		+ \tfrac12 d_{\mathbf {14} ij}^2
	-\tfrac1{18} |f_{\mathbf 1 ijk}+\tfrac{i}2d_XF_{ijk}|^2
		-\tfrac1{24} |f_{\mathbf 7 ijk}|^2
		+\tfrac1{24} |f_{\mathbf {27} ijk}|^2
.
\end{align}
Here we used (\ref{E:H3form}) and the fact that for any 2-form $\eta$,
\begin{align}
\eta_{ij}\psi^{ijkl} \eta_{kl} &= -4 \eta_{\mathbf7 ij}^2 
	+ 2 \eta_{\mathbf{14} ij}^2
\end{align}
as follows from (\ref{E:H2form}).
Substituting the algebraic equations (\ref{E:AuxiliaryEOM}) back into the component action and using (\ref{E:contractions}), we find the following scalar potential:
\begin{align}
\label{E:ScalarPotential}
V(x,y)&= 
	-\tfrac{42}{18}\cdot \tfrac 9{16}\tau_0^2 
	- \left(6+\tfrac{9\cdot 24}{24}\right)\tau_1^2 
	+\tfrac12\cdot\tfrac14 \tau_2^2 
	+\tfrac1{24} \tau_3^2			
\cr&
	+\left(\tfrac{49}{16} - \tfrac{42}{18} \cdot \tfrac 9{16}\right) \sigma_0^2 
	+\left(2\cdot 9 - \tfrac{9\cdot 24}{24} \right) \sigma_1^2  
	+\tfrac1{24} \sigma_3^2		
\cr&= 
	-\tfrac{21}{16} \tau_0^2 
	- 15\tau_1^2 
	+\tfrac18 \tau_2^2 
	+\tfrac1{24} \tau_3^2			
	+\tfrac{7}{4} \sigma_0^2 
	+ 9 \sigma_1^2  
	+\tfrac1{24} \sigma_3^2		
\end{align}
To interpret this, we appeal to a result of Bryant who has computed that the scalar curvature of the metric is given by \cite{Bryant2003}
\begin{align}
\label{E:Bryant}
R(g) &=	12 \nabla^i (\tau_1)_i
			+\tfrac{21}{8} \tau_0^2 
			+ 30\tau_1^2 
			-\tfrac14 \tau_2^2 
			-\tfrac1{12} \tau_3^2			
\end{align}
when written in terms of the intrinsic torsion forms (\ref{E:TorsionClassesDef}).
Similarly, one checks that the $\sigma_\mu$ terms combine into a Maxwell term.
Therefore, up to a surface term, we have shown that 
\begin{align}
\label{E:Bryant}
-V(x,y)&=	
		\tfrac12 R(g)
		-\tfrac1{4\cdot 4!} F_{ijkl}^2 
,
\end{align}
which is the potential of the bosonic part of eleven-dimensional supergravity. 
We conclude that the action (\ref{E:GravitinoExpansion}) reproduces the correct scalar potential of eleven-dimensional supergravity.

\section{Discussion}
\label{S:Discussion}

In this paper we computed the potential of eleven-dimensional supergravity as it is described in a superspace background of the form $\mathbf R^{4|4} \times Y$ with $Y$ a (not necessarily compact) manifold with $G_2$ structure. 
To first order in a gravitino superfield expansion (\ref{E:GravitinoExpansion}), the action is the sum of two terms. 
The first is the superspace Chern-Simons invariant (\ref{E:NACS}) of the gauged tensor hierarchy of the eleven-dimensional 3-form.
This tensor hierarchy is a $\mathfrak g$-equivariant superspace chain complex with $\mathfrak g$ the algebra of diffeomorphisms on $Y$ \cite{Becker:2016xgv, Becker:2016rku}. 

The second is a superspace version of the Hitchin functional for $G_2$-structure manifolds (\ref{E:Kahler}).
The bosonic version of this functional is the volume of $Y$ as computed from the Riemannian metric 
constructed from an arbitrary (stable) 3-form.
(The stationary points of this functional on cohomology classes of the 3-form define $G_2$-holonomy metrics.) 
This functional is lifted to superspace by formally replacing the 3-form with the tensor hierarchy superfield strength containing the gravitational scalars and integrating over superspace. 
(The 3-form on $Y$ on which the tensor hierarchy is based is embedded as the imaginary part of a chiral superfield; the stable 3-form is the real part.)

Having defined the action thus, we computed its potential by integrating out the auxiliary fields of the multiplets in the tensor hierarchy. 
What we find is the Einstein-Hilbert action on $Y$ in the form computed by Bryant \cite{Bryant2003} with an analogous form for the Maxwell term for the 3-form scalars.
As this is the correct potential for eleven-dimensional supergravity on $X \times Y$, this observation relates topological M theory \cite{Dijkgraaf:2004te} to ``physical'' M-theory, and suggests the following construction of eleven-dimensional supergravity on $\bm X\times Y$ to this order in the super-gravitino expansion (\ref{E:GravitinoExpansion}): 
Starting with the non-abelian tensor hierarchy, one defines on the space of 3-form field strength superfields the curved superspace generalization (\ref{E:Kahler}) of the Hitchin functional. To this one adds the Chern-Simons super-invariant (\ref{E:NACS}) in curved superspace \cite{Randall:2016zpo} (see also \cite{Aoki:2016rfz, Yokokura:2016xcf}). 
Note that we cannot add a superpotential beyond the F-terms coming from the Chern-Simons action because the $G_2$-structure metric is not chiral. 
Therefore, we might expect this to be the full answer at this order of the gravitino expansion. 

As it stands, this proposal probably requires some modification, or at least a better understanding of the following puzzle: 
In freezing the supergravity fields (\ref{E:Freezing}) and ignoring the superfields containing the gauge 2-forms, we are implicitly assuming that these fields do not carry propagating scalars. 
In particular, one needs to explain the mechanism by which the superfluous scalars in the 3- and 2-form multiplets are removed from the spectrum. 

A potential resolution to this problem is that the 4-form field strength actually {\em is} the supergravity conformal compensator $G=\Phi_0^3$. Similarly the 3-form field strengths $H_i$ would be compensators for the extended $R$-symmetry. In fact, the Hitchin metric is negative definite on these representations (cf.\ eq.\ \ref{E:H3form}) which would result in the wrong-sign kinetic terms for these fields that is the hallmark of a compensating field \cite{Gates:1983nr}.
In such a scenario the super-diffeomorphisms of the theory would allow one to fix $G$ and $H_i$ by a choice of conformal and extended $R$-symmetry gauge. 

In this interpretation, the $\int G \Phi \partial \Phi$ term in the Chern-Simons action (\ref{E:NACS}) becomes
\begin{align}
-\frac i{12 \kappa^2} \int d^4x  \int_Y  \int d^2\theta\, \Phi_0^3 \,  \Phi \wedge \partial \Phi + \mathrm{h.c.}
.
\end{align}
This is the gravitational covariantization of the chiral superpotential postulated in reference \cite{Becker:2014rea}
in the old-minimal formulation of 4D, $N=1$ supergravity \cite{Wess:1977fn, Wess:1978bu}.
This superpotential combines the action constructed by Gukov \cite{Gukov:1999gr} (see also \cite{Gukov:1999ya}) in the context of flux compactifications with the terms needed for holomorphicity in general backgrounds. 

We can extend this interpretation to the K\"ahler term as follows by covariantizing of the Hitchin functional \cite{Gates:1983nr}
\begin{align}
\label{E:GravKahler}
\int d^4 x \int d^7y \int d^4\theta \,E  \,\bar \Phi_0\Phi_0 \, \sqrt{g(F)} 
.
\end{align}
Under the identification of the compensator as the cube root of $G$, we recognize this as the (non-mass part of the) K\"ahler action (\ref{E:Kahler}). 

As dicussed in references \cite{Acharya:2000ps, Beasley:2002db, House:2004pm}, the original flux potential gets corrections from M2-brane domain walls localized at points on $Y$ and/or M5-branes wrapped on associative 3-cycles in $Y$. 
There it is argued that these corrections change the superpotential by what is essentially the current of the Page charge \cite{Page:1984qv}. They then write the corrected Gukov superpotential as $W\sim m \mathrm{vol}(Y)  + \int_Y  \Phi \wedge d \Phi$ where $m$ is the Freund-Rubin mass \cite{Freund:1980xh}.
Going back to superspace, the contribution of the first term would come from an expression of the form (\ref{E:GravKahler}).

Finally, we mention that without including the (spacetime scalar) auxiliary fields of all of the superfields in the theory, one does not expect to recover the correct scalar potential. Since 4D, $N=1$ supergravity contains two such fields, these should have already been included in our analysis lest they over-correct the potential. In the identification above, these fields were already taken into account by identifying the supergravity scalar auxiliary with $d_X$ and the pseudo-scalar with the dual of the 4-form field strength $F_{abcd} = 4\partial_{[a} C_{bcd]}$. 
An analogous resolution of the puzzle for the scalars in $H_i$ will have to await (and hint at) the construction of the couplings to the gravitino multiplets. 
These couplings are currently under investigation.

\section*{Acknowledgements}
It is a pleasure to thank Daniel Butter and Stephen Randall for enlightening discussions and Sergei Gukov and Nikita Nekrasov for comments and suggestions.
W{\sc dl}3 and D{\sc r} are grateful to the Simons Center for Geometry and Physics for hospitality during the {\sc ix} Simons Summer Workshop. 
This work is partially supported by NSF Focused Research Grant DMS-1159404 and the Mitchell Institute for Physics and Astronomy at Texas A\&M University.

\appendix
\section{$G_2$ Toolbox}
\label{S:G2}
In this appendix, we review the construction of the Hitchin functional for $G_2$-structure 7-manifolds and collect various identities of $G_2$ linear algebra \cite{joyce2000compact, Bryant2003, Karigiannis:0301218}.
Let $\varphi$ be a 3-form on $Y$ and define the symmetric bilinear form 
\begin{align}
s_{ij}  :=  -\tfrac1{144} \epsilon^{abcdefg}\varphi_{iab} \varphi_{cde} \varphi_{jfg}
.
\end{align}
The 3-form $\varphi$ is ``stable'' iff $\mathrm{det}(s)\neq 0$.\footnote{Stability as formulated in \cite{Hitchin:2000jd, Hitchin2001} is in terms of open orbits of the $GL(n)$ action on the space of $p$-forms on the tangent bundle of an $n$-manifold $Y$. This condition is the precise criterion for when a volume form constructed from fractional powers of the $p$-form exists.  In order for $g_{ij}$ to be a good metric, we actually need that $\varphi$ is positive, implying that $s_{ij}$ and $g_{ij}$ are positive definite, but this is only a slightly stronger condition, since if $\varphi$ is stable than either $\varphi$ or $-\varphi$ is positive. In this paper we will simply take stability to mean $\mathrm{det}(s)\neq 0$, and we will not always emphasize positivity.
} 
We assume this non-degeneracy condition throughout the paper. 
A stable 3-form on the tangent spaces of $Y$ reduces the structure group $GL(7) \to G_2$. 
Thus, our assumption implies that $Y$ is a $G_2$-structure manifold. 

Normalizing
\begin{align}
\label{E:metric}
g_{ij} = s^{-1/9} s_{ij} 
~~~\Leftrightarrow~~~
\sqrt{g} g_{ij} = s_{ij},
\end{align}
defines the Riemannian metric $g$ on $Y$. 
We can construct the Riemannian volume functional from the determinant of the metric
\begin{align}
\label{E:HitchinFunctional}
\Phi(\varphi) := \int_Y d^{11}y \, \sqrt{g(\varphi)}
\end{align}
This expression is (equivalent to) the Hitchin functional on the space of stable 3-forms on $Y$ \cite{Hitchin:2000jd}. 
In that reference, it is shown if $(Y,\varphi)$ a closed $G_2$-holonomy manifold, then $\varphi$ is (closed by definition and) a critical point of $\Phi$ restricted to the cohomology class $[\varphi]\in H^3(Y, \mathbf R)$. 
Conversely, if $\varphi$ is a critical point on a cohomology class of a closed oriented 7-manifold $Y$ such that $\varphi$ is stable, then $\varphi$ defines a metric with $G_2$ holonomy. 
For any $p$-form $\omega$, let $\omega_\mathbf i := \pi_i \omega$ denote the projection to the $\mathbf{i}$-dimensional representation. Then, under a variation $\delta \varphi$ of the $G_2$ structure form 
\begin{align}
\label{E:No7}
\delta g_{ij} = \varphi_{(i}{}^{kl} \left[ 
	\tfrac19 (\delta\varphi)_\mathbf1
	+\tfrac12 (\delta\varphi)_\mathbf{27}
	\right]_{j)kl}
,	
\end{align}
the metric does not transform (to first order) under the $\mathbf 7$ projection of the variation.
We will not need these facts for this paper; we include them only to motivate the definition of the Hitchin functional. 
(To the interested reader, we recommend Karigiannis' thesis \cite{Karigiannis:0301218}.)

We now review some $G_2$ linear algebra and define the projectors from representations of $SO(7)$ to those of $G_2$.
Under the reduction $SO(7) \to G_2$ of the structure group, the $\mathbf{21}$-dimensional space of 2-forms on $Y$ decomposes into $G_2$ representations as $\mathbf{21}=\mathbf{7}\oplus \mathbf{14}$. 
Similarly, the $\mathbf{35}$-dimensional space of 3-forms on $Y$ decomposes as $\mathbf{35}=\mathbf{1}\oplus\mathbf{7}\oplus \mathbf{27}$. 
(We review the explicit formul\ae{} for the projectors to these representations presently.) 
We start by defining the dual $\psi = \ast \varphi$ with components
\begin{align}
\psi_{ijkl} = \tfrac1{3!} \sqrt{g} \epsilon_{ijklmnp}g^{mm'}g^{nn'}g^{pp'} \varphi_{m'n'p'}
\end{align}
(the opposite is $\varphi_{ijk} = \tfrac1{4!} \sqrt{g} \epsilon_{ijklmnp}g^{mm'}g^{nn'}g^{pp'}g^{qq'} \psi_{m'n'p'q'}$).
Useful identities include
\begin{align}
\label{E:contractions}
\varphi^{ijk} \varphi_{ij'k'} = 2\delta_{[j'}^j\delta_{k']}^k - \psi_{j'k'}{}^{jk}
~~,~~~
\psi^{ijkl} \psi_{ijk'l'} = 8 \delta_{[k'}^k \delta_{l']}^l - 2 \psi_{k'l'}{}^{kl}
~~,
\cr
\varphi^{ijk} \varphi_{ijk'} = 6\delta_{k'}^k
~~,~~~
\psi^{ijkl} \psi_{ijkl'} = 24 \delta_{l'}^l
~~,~~~
\varphi_i{}^{lm} \psi_{jk lm} = - 4 \varphi_{ijk} ~,
\end{align}
where indices are raised and lowered with the metric (\ref{E:metric}).
These identities can be used to construct the projectors from the representations of $SO(7)$ onto the irreducible representations of $G_2$: For any 2-form $\eta$ and 3-form $\omega$,
\begin{subequations}
\begin{align}
\pi_7 \eta_{ij} &= \left( \tfrac13\delta_i^k \delta_j^l  - \tfrac16\psi_{ij}{}^{kl} \right) \eta_{kl}
\\
\pi_{14} \eta_{ij} &= \left( \tfrac23\delta_i^k \delta_j^l  +\tfrac16 \psi_{ij}{}^{kl} \right) \eta_{kl}
\\
\label{E:35to1}
\pi_1 \omega_{ijk} &= \tfrac1{42} \varphi_{ijk} \varphi^{i'j'k'} \omega_{i'j'k'}
\\
\label{E:35to7}
\pi_7 \omega_{ijk} &=  \left(
	\tfrac14\delta_i^{i'}\delta_j^{j'}\delta_k^{k'} 
	-\tfrac38 \psi_{[ij}{}^{i'j'}\delta_{k]}^{k'} 
	-\tfrac1{24}\varphi_{ijk} \varphi^{i'j'k'}\right) \omega_{i'j'k'}
\\
\pi_{27} \omega_{ijk} &=  \left(
	\tfrac34\delta_i^{i'}\delta_j^{j'}\delta_k^{k'} 
	+\tfrac38 \psi_{[ij}{}^{i'j'}\delta_{k]}^{k'}  
	+\tfrac1{56}\varphi_{ijk} \varphi^{i'j'k'}\right) \omega_{i'j'k'}
.	
\end{align}
\end{subequations}
Two useful identities on the space of 2-forms are
\begin{align}
\label{E:H2form}
\psi_{ij}{}^{kl} \eta_{\mathbf 7 kl} = -4 \eta_{\mathbf 7 kl}
~~~,~~~
\psi_{ij}{}^{kl} \eta_{\mathbf {14} kl} = 2 \eta_{\mathbf {14} kl}
\end{align}
Similarly, on the space of 3-forms, 
\begin{subequations}
\label{E:3formSquares}
\begin{align}
\omega^2 := \omega^{ijk}\omega_{ijk} &= 
	\omega_\mathbf 1^2 + \omega_\mathbf 7^2 + \omega_{\mathbf {27}}^2 
\\
g^{ii'} \psi^{jk j'k'} \omega_{ijk}\omega_{i'j'k'} &= 
	- 4 \omega_\mathbf 1^2 -2 \omega_\mathbf 7^2 + \tfrac23 \omega_{\mathbf {27}}^2 
\\
\label{E:42}
(\varphi^{ijk}\omega_{ijk})^2 &= 
	42 \omega_\mathbf 1^2
.
\end{align}
\end{subequations}

{
\footnotesize
\bibliography{/Users/wdlinch3/Dropbox/Rashoumon/LaTeX/BibTex/BibTex}
\bibliographystyle{unsrt}
}

\end{document}